\documentclass[aps,prd,twocolumn,superscriptaddress]{revtex4}

\usepackage{graphicx}         
\usepackage{amssymb}
\usepackage{amsmath}
\usepackage{amsthm}
\usepackage[colorlinks]{hyperref}
\usepackage{color,rotating}
\usepackage{bbm}
\usepackage{array}
\usepackage{pstricks}
\usepackage{pstricks-add}
\usepackage{colonequals}
\usepackage{cleveref}
\usepackage{bbold}
\usepackage{verbatim}

\definecolor{darkgreen}{rgb}{0,0.6,0}
\definecolor{gray}{rgb}{.7,.7,.7}

\DeclareMathAlphabet{\EuRoman}{U}{eur}{m}{n}
\SetMathAlphabet{\EuRoman}{bold}{U}{eur}{b}{n}

 \graphicspath{{./figures/}}

\def\di{\displaystyle}

\def\bg{\begin{eqnarray}\begin{array}{rcl}\displaystyle}
\def\eg{\end{array} &\di    &\di   \end{eqnarray}}
\def\bm#1{\begin{eqnarray}\begin{array}{#1}\di}
\def\bmo#1{\begin{eqnarray*}\begin{array}{#1}\di}
\def\bml#1#2{\begin{eqnarray}\begin{array}{#1}\label{#2}\di}
\def\bgo{\begin{eqnarray*}\begin{array}{rcl}\displaystyle}
\def\ego{\end{array} &\di    &\di \nonumber  \end{eqnarray*}}

\def\btensor#1#2{\renew\left#1\begin{array}{#2}\di}
\def\brtensor#1#2#3{\ren#3\left#1\begin{array}{#2}}
\def\botensor#1#2{\renew\left#1\begin{array}{#2}}
\def\etensor#1{\end{array}\right#1}

\def\s0#1#2{\mbox{\small{$ \frac{#1}{#2} $}}}
\def\0#1#2{\frac{#1}{#2}}

\def\A{A\!\llap{/}}

\def\s{\sigma}

\def\ren#1{\renewcommand{\arraystretch}{#1}}

\def\renew{\renewcommand{\arraystretch}{1}}

\begin{document}

\title{Four-Derivative Quantum Gravity Beyond Perturbation Theory}

%
\author{N. Christiansen}
\affiliation{Institut f\"ur Theoretische Physik, Universit\"at Heidelberg,
Philosophenweg 16, 69120 Heidelberg, Germany}
\email{n.christiansen@thphys.uni-heidelberg.de}


\begin{abstract}
In this work we investigate the ultraviolet behavior of Euclidean 
four-derivative quantum gravity beyond perturbation theory. In addition to 
a perturbative fixed point, we find an ultraviolet fixed point that is 
non-trivial in all couplings and is described by only two free 
parameters. This result is in line with the asymptotic safety scenario in 
quantum gravity. In particular, it supports the conjecture that the full 
theory is described by a finite number of free parameters.
\end{abstract}

\maketitle

\section{Introduction}

The quantum description of spacetime is one of the fundamental open questions 
in theoretical physics. An approach that satisfies the principle of minimal 
assumptions about unknown physics is the asymptotic safety scenario, which 
was introduced by Steven Weinberg in 1976, \cite{Weinberg:1980gg}. The 
underlying idea is that the renormalization group flow of quantum gravity 
approaches a non-trivial, i.e., non-Gaussian, ultraviolet fixed point with a 
finite dimensional UV-critical 
surface. The first property ensures finiteness, while the latter makes the 
theory predictive in the sense that it contains only a finite number of free 
parameters. It is well-known that the standard perturbative renormalization 
programme for general relativity fails: the theory turns out to be 
perturbatively non-renormalizable \cite{tHooft:1974toh,Goroff:1985sz}. Weinbergs 
idea, however, is not limited to 
perturbative quantization of general relativity, but provides the possibility 
for a non-perturbative UV-completion of the theory. Nevertheless, the theory is 
described by a quantum field theory where it is only assumed that 
there is an underlying diffeomorphism symmetry and that the degrees of freedom 
are carried by a spin-two field. Therefore, there are no further assumptions 
about unknown new physics at some high energy scale.
\\The explicit framework to search for asymptotic safety in quantum 
gravity was set up by Reuter \cite{Reuter:1996cp} based on the  
functional renormalization group and the Wetterich equation, 
\cite{Wetterich:1992yh}. 
During the last two decades, numerous publications investigated the ultraviolet 
behavior of the renormalization group flow of quantum gravity and compelling 
hints for the existence of a non-Gaussian fixed point were found,
\cite{Reuter:2001ag,Lauscher:2002sq,Litim:2003vp,Fischer:2006fz,
Codello:2006in,Machado:2007ea,Codello:2007bd,Benedetti:2009rx,Eichhorn:2009ah,
Manrique:2009uh,Niedermaier:2009zz, 
Eichhorn:2010tb,Groh:2010ta,Manrique:2010am,Manrique:2011jc,Christiansen:2012rx, 
Christiansen:2014raa,Christiansen:2015rva,Donkin:2012ud,Benedetti:2012dx,
Dietz:2012ic,Nink:2012vd, 
Becker:2014qya,Falls:2013bv,Ohta:2013uca,Benedetti:2013jk,Falls:2014tra, 
Eichhorn:2015bna,Falls:2015qga,Falls:2015cta,Ohta:2015fcu,Ohta:2015efa, 
Falls:2016msz,Falls:2016wsa,Gies:2016con,Gies:2015tca,Labus:2016lkh,
Biemans:2016rvp}. 
This also 
extends to theories where gravity is coupled to matter and gauge fields, 
\cite{Dou:1997fg,Percacci:2002ie,Narain:2009fy,Daum:2009dn,Eichhorn:2011pc,
Folkerts:2011jz, 
Harst:2011zx,Dona:2013qba,Henz:2013oxa,Meibohm:2015twa,
Oda:2015sma,Henz:2016aoh,Meibohm:2016mkp,Eichhorn:2016esv,Eichhorn:2016vvy}. For 
reviews on asymptotic safety in quantum 
gravity see 
\cite{Niedermaier:2006wt,Niedermaier:2006ns,Percacci:2007sz,
Litim:2008tt,Codello:2008vh,
Litim:2011cp,Reuter:2012id,Nagy:2012ef,Reuter:2012xf,Ashtekar:2014kba}. The 
general 
idea of asymptotic safety is of course not restricted to quantum 
gravity, but non-Gaussian fixed points are of general interest in quantum field 
theory. Asymptotically 
safe theories without gravity are studied e.g.\ in 
\cite{Gawedzki:1985jn,Gawedzki:1985ed,Gies:2009hq,Gies:2009sv,Braun:2010tt,
Gies:2013pma,
Litim:2014uca, 
Esbensen:2015cjw,Boettcher:2015pja,Eichhorn:2016hdi,Bond:2016dvk}.   
\\A particularly interesting type of gravity theories are 
four-derivative theories, i.e.\ theories with an action that includes not only 
Einstein-Hilbert terms but also operators with mass dimension four. The 
couplings of these operators are dimensionless in four spacetime dimensions and 
the propagator has a 
$1/p^4$ falloff, which makes the theory perturbatively renormalizable, 
\cite{PhysRevD.16.953}. Using a complete basis of four-derivative operators, 
it was shown explicitly in one-loop calculations that the theory is 
asymptotically free in the coupling of the Weyl-squared tensor, while it 
exhibits a non-trivial UV-fixed point in the other 
couplings, \cite{Fradkin:1981hx,Avramidi:1985ki}. Such theories were also 
studied in the context of the asymptotic safety scenario, 
\cite{Codello:2006in,Benedetti:2009rx,Niedermaier:2009zz,Niedermaier:2010zz}.
In \cite{Benedetti:2009rx}, also a purely non-Gaussian
fixed point was discovered in the four-derivative theory. Despite being 
perturbatively 
renormalizable, higher derivative theories did not receive a lot of attention 
as candidates for a fundamental quantum theory of gravitation during the last 
decades. The 
reason is that squares of the Ricci tensor induce an additional pole with 
negative residue in the graviton propagator around flat background. It is 
believed that this feature 
spoils unitarity of the resulting quantum field theory. However, one can offer 
several objections against this claim. First of all, the faith of this pole in 
a resummed graviton propagators is not clear, and second, the non-perturbative 
relation between poles in the Euclidean propagator and unitarity of the theory 
in real-time, i.e.\ the spectral reconstruction, is highly non-trivial. Hence, 
the 
question of unitarity should be considered as an open issue, but cannot be 
used to abandon four-derivative theories right away.        
\\In this work we address several aspects of the asymptotic safety scenario in 
quantum gravity. We 
use the formalism for vertex expansions developed in 
\cite{Christiansen:2014raa}. In this approach the effective 
action is expanded 
around a flat background in powers of the graviton field. The latter drives the 
renormalization group flow and can be disentangled from the background field 
and the related couplings. In this formalism we 
include for the first time tensor structures beyond Einstein-Hilbert, and use a 
complete basis of diffeomorphism invariant four-derivative operators. This is 
then used to calculate the scale-dependence of the graviton propagator in a 
non-perturbative fashion from which we extract the scale-dependence of the 
graviton wave-function renormalization and the running couplings associated 
with the volume term, the $R$, $R^2$ and $R_{\mu \nu}^2$ operators. In order to 
derive the corresponding beta-functions we present several technical advances 
in the context of flow equations. 
\\The non-Gaussian fixed point in the Einstein-Hilbert truncation, where the 
set of $\beta$ functions is determined by the gravitational constant $G$ and 
the cosmological constant $\Lambda$, is characterized by two relevant 
directions. This implies that the theory is described by two free parameters.
As predictivity is encoded in a finite number of relevant directions, the 
asymptotic safety conjecture heavily relies on the identification of a pattern 
that guarantees irrelevance of higher order operators. In this work we find 
a non-Gaussian fixed point in four dimensional coupling space that 
exhibits only two relevant directions. Thus, the theory has not only a 
well-defined ultraviolet limit, but the classically marginal four-derivative 
operators do not induce further relevant directions. This is a very encouraging 
structure, as it suggests that quantum fluctuations do not turn irrelevant into 
relevant operators and that the UV-critical surface is indeed finite 
dimensional. 
\\On the technical side, we generalize the setup 
in \cite{Christiansen:2014raa,Christiansen:2015rva} 
in the presence of higher order operators, including the projection on the 
coupling constants. Moreover, we use a gauge fixing condition that is different 
from the usual choice used in four-derivative gravity, but is 
found to be the natural choice in the present setup. Indeed, it has been 
shown in \cite{Wetterich:2016ewc} that this particular choice for the 
gauge-fixing functional induces a decoupling of gauge fluctuations.

\section{Functional Renormalization and Vertex Construction}

\subsection{The Wetterich Equation and Vertex Flow Equations}

\begin{figure}
\begin{align*}
\partial_t \left. \frac{\delta^2\Gamma_k[\bar{g};h]}{\delta h^2}
\right|_{h=0} &= -\frac{1}{2} \includegraphics[height=6ex]{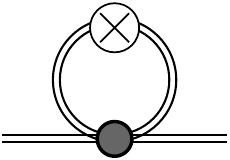} +
\raisebox{-2ex}{\includegraphics[height=6ex]{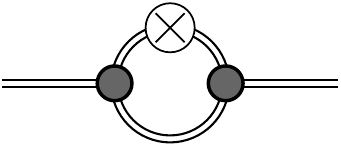}}\\
	      & \mathrel{\hphantom{=}} - 2
\,\raisebox{-2.2ex}{\includegraphics[height=6ex]{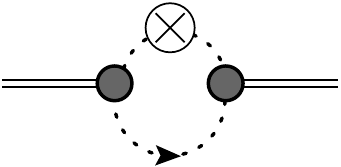}} \equiv
\mathrm{Flow}^{(2)}
\end{align*}
\caption{Diagrammatic representation of the flow equation for the second order 
vertex function $\Gamma^{(2)}$. In contrast to the usual perturbative 
Feynman-diagrams, which are build from tree-level quantities, the propagators 
and the vertices in the flow equations are fully dressed, i.e.\ they contain 
all possible quantum fluctuations up the scale $k$.  
The dressed graviton propagator is represented by the double line, the
ghost  propagator by the dashed line, while a dressed vertex is denoted by a
dot. The regulator insertion is indicated by the crossed circle.}
\label{fig:flow_of_prop}
\end{figure}

The basic ingredients for an investigation of the asymptotic safety conjecture 
are the $\beta$-functions beyond standard perturbative expansions. The 
functional renormalization group is a non-perturbative approach to continuum 
quantum field theory and in particular its formulation for the 1PI effective 
action with the Wetterich equation \cite{Wetterich:1992yh} has proven to be a 
very powerful method. It is based on the Wilsonian idea of coarse graining
by successively integrating out infinitesimal momentum shells.
This idea is implemented with a regulator term in the path integral, which 
introduces a cutoff-scale $k$. This finally leads to a functional differential 
equation that determines the scale-dependence of the quantum effective action 
$\Gamma_k$, which now depends on the RG-scale $k$. 
An additional complication in gravity is that 
this regulator necessitates the introduction of a background field $\bar{g}$. 
Besides technical reasons, a background metric $\bar{g}$ in the 
regulator is needed in order to construct a differential operator that defines 
via its spectrum the meaning of large and small momenta. 
In addition to that, the Wetterich equation is formulated in terms of 
propagators. Therefore 
one needs to work with a gauge fixed theory, which in turn requires a 
background field. As a result, the quantum 
effective action $\Gamma[\bar{g},\phi]$ depends on the background $\bar{g}$  
and and a fluctuation super-field $\phi =(h,\bar{c},c)$. In 
gravity this super-field contains the dynamical graviton field $h$, as well as 
the Faddeev-Popov ghost fields $\bar{c}$ and $c$. With these ingredients 
the Wetterich equation for quantum gravity reads        
\begin{align}\label{floweq}
\notag \partial_t \Gamma_k[\bar{g},\phi] = \frac{1}{2} & \mathrm{Tr} \,
\left[\left(\Gamma^{(2h)}_{k} + R_{k,h} \right)^{-1} \,\partial_t R_{k,h}
\right][\bar{g},\phi]
\\[2ex]  - & \mathrm{Tr} \, \left[ \left(\Gamma^{(\bar{c}c)}_{k} + R_{k,c} 
\right)^{-1}\,\partial_t
R_{k,c}\right][\bar{g},\phi] \, ,
\end{align}
and we use the abbreviation
\begin{equation}
\Gamma^{(\phi_1 ... \phi_n)}_k[\bar{g},\phi] \colonequals 
\frac{\delta^n \Gamma_k[\bar{g},\phi]}{\delta \phi_1 \cdots \delta
\phi_n} \, 
\end{equation}
for functional derivatives. In the above functional differential equation $R_k$ 
is the regulator function that organizes local momentum integration of 
fluctuations with $q \approx k$. Moreover, $t$ is the logarithmic RG-scale $t := 
\log (k/k_0)$ with an arbitrary reference scale $k_0$. The Tr in the above flow 
equation denotes a summation over all discrete indices and an integration over 
continuous ones. We will also use the notation $\partial_t f(k) =: \dot{f}(k)$ 
for any function $f$.  
\\There are several important issues concerning the role the background field 
in 
the 
flow equation. According to the general principles of gravity, physical 
observables should be independent of an auxiliary background $\bar{g}$ that 
needs to be introduced for technical reasons. In the present formalism with the 
two fields $\bar{g}$ and $\phi$, the effective action $\Gamma[\bar{g},\phi]$ 
is truly a functional of two fields, and the dependence cannot be combined into 
a 
single, physical metric $g = \bar{g} + \phi$. In particular, it follows that 
the vertex functions $\Gamma^{(n)}$ are explicitly background dependent. 
However, these correlation functions are not directly related to observables 
and their explicit background dependence is indeed necessary in order to 
guarantee background independence of physical observables. The separate 
dependence on the two fields $\bar{g}$ and $\phi$ is encoded in non-trivial 
Nielsen identities, also called split-Ward-identities 
\cite{Litim:2002ce,Pawlowski:2003sk,Pawlowski:2005xe,Donkin:2012ud,
Becker:2014qya,Dietz:2015owa,Labus:2016lkh,Wetterich:2016ewc,
Morris:2016spn,Percacci:2016arh}. In the standard 
background field approximation one evaluates the Wetterich equation 
\eqref{floweq} at vanishing fluctuation field $\phi=0$. However, this does 
not lead to a closed equation as 
\begin{equation}
\left.\frac{\delta^2 \Gamma_k[\bar{g},\phi]}{\delta h^2}\right|_{\phi=0} \neq
\frac{\delta^2 \Gamma_k[\bar{g},0]}{\delta \bar{g}^2} \, .
\end{equation} 
Consequently, by using this approximation one does not calculate correlation 
functions of the fluctuation field $\phi$, but correlations of the background 
field $\bar{g}$. In order to circumnavigate this problem, we can make use of 
the infinite hierarchy of flow equations that is generated by the master 
equation \eqref{floweq}. This hierarchy is obtained by taking functional 
derivatives of the Wetterich equation,
\begin{equation}\label{flow_hierarchy}
\frac{\delta^n}{\delta \phi^n} \dot{\Gamma}_k[\bar{g},h] = 
\mathrm{Flow}^{(n)}[\Gamma^{(2)},...,\Gamma^{(n+2)}] \,, 
\end{equation}
where $\mathrm{Flow}^{(n)}$ denotes the $n$-th functional derivative of the RHS 
of \eqref{floweq}. It is important to note that the flow of the vertex function 
of order $n$ depends on the vertex functions of order two up to order $n+2$.
 With these relations we are equipped with equations for the $n$-th 
moments of the effective action, which define the quantum field 
theory. This approach has the additional advantage that one 
gains access to the momentum dependence of the vertex functions $\Gamma^{(n)}$ 
and that their dependence on the $RG$-scale $k$ can be studied separately.
In this work we will study the flow equation for the inverse propagator 
$\Gamma^{(2)}$, which has the diagrammatic representation shown in 
\autoref{fig:flow_of_prop}. 

\subsection{Vertex Functions}

In this section we turn to the construction of the vertex functions, which are 
the essential building blocks in the flow equation. The general setup is based 
on the formalism presented in \cite{Christiansen:2014raa}, but is generalized 
to tensor structures beyond Einstein-Hilbert.
Our goal is to construct an approximation based on a vertex expansion, i.e.\ a 
functional Taylor expansion of the effective action in powers of the 
fluctuation field $h$ according to
\begin{equation}\label{vertex_exp}
\Gamma[\bar{g},h] =\sum_{n=0}^{N} \frac{1}{n!} \left. 
\frac{\delta^n \Gamma[\bar{g},h]}{\delta h^n} \right|_{h=0} h^n \,.
\end{equation}
The above notation is symbolic and indices and its contractions as well 
as spacetime integrals are suppressed
In this work we choose flat spacetime as the expansion point of the effective 
action, i.e.,  $\bar{g}_{\mu \nu} = \mathbb{1}$.
The most general form of the vertex functions 
is not unique, but one can choose different paramterizations. 
A canonical form is given by
\begin{equation}
\Gamma^{(n)}_k(p_1,...,p_n) = \sum_i g^{(n)}_i(k,p_1,...,p_n) 
\mathcal{T}^{(n)}_i(p_1,...,p_n) \, , 
\end{equation}
where $\mathcal{T}^{(n)}_i$ are tensor structures that form a basis in the 
relevant tensor space. The $g^{(n)}_i(k,p_1,...,p_n)$ are parameters, which 
in general depend not only on the RG-scale $k$, but also on all 
external momenta $p_1,...,p_n$. However, this is obviously far too general 
for practical computations. In order to construct approximations to this 
most general form, there are several guiding principles that underlie the 
following construction of vertex functions. First of all, it is important to 
note that the expansion \eqref{vertex_exp} is not diffeomorphism invariant nor 
background independent. In particular, the vertex functions $\Gamma^{(n)}$ 
inherit this property. Nevertheless, we want to restrict their tensor 
structures to the ones that originate from functional derivatives of 
diffeomorphism invariant operators. This is motivated by the conjecture that 
diffeomorphism invariance is broken only weakly, which is observed in 
\cite{Christiansen:2015rva} and \cite{4_pt}. This also what is expected in 
semi-perturbative regimes.  
In the present work, we are 
interested in four-derivative gravity, and thus our tensor structures are 
generated by the action    
\begin{align}\label{gen_action}
S_{G,\Lambda,a,b}[g] = \frac{1}{16 \pi} \int_x 
\left(\frac{2\Lambda}{G} -\frac{R}{G} 
+ a \, R^2 + b \, R^2_{\mu \nu}\right) \, ,    
\end{align}
where we have defined $\int_x := \int \mathrm{d}^4x \sqrt{\mathrm{det}g}$. 
Moreover, we 
work in Euclidean spacetime throughout this work. 
Action \eqref{gen_action} is the most general diffeomorphism invariant action 
that contains 
up to four derivatives of the metric. More precisely, the operators $R^2$ 
and $\mathrm{Ricc}^2$ are a basis of local diffeomorphism invariants in four 
spacetime dimensions if we drop boundary terms, i.e.\ terms that are total 
derivatives and do not contribute to local physics. Hence, we ignore a 
$\Delta R$ term in the action. Moreover, the Riemann-tensor squared 
$\mathrm{Riem}^2$ can be written as a linear combination of $R^2$, 
$\mathrm{Ricc}^2$ and a 
topological invariant due to the generalized Gauss-Bonnet theorem in four 
spacetime dimensions. Additionally, the $\mathrm{Ricc}^2$ term can always be 
traded for the square of the Weyl-tensor $C^2$. In the basis with the $C^2$ 
operator, the most common paramterization reads 
 \begin{equation}\label{gen_action_reparam}
S_{G,\Lambda,\omega,s}[g] = \frac{1}{16 \pi} \int_x
\left(\frac{2\Lambda}{G} -\frac{R}{G} 
+ \frac{\omega}{s} \, R^2 + \frac{1}{s} \, C^2\right) \,,
\end{equation}
where the four-derivative operators have the common coupling $1/s$ and the 
relative interaction strength of $R^2$ and $C^2$ is encoded in the coupling 
$\omega$. The different couplings in \eqref{gen_action} and 
\eqref{gen_action_reparam} are related by simple algebraic equations.
Most of the time we will use the parameterization \eqref{gen_action}.  

The vertex construction used in this paper is based on a 
quantum deformation of the classical vertices, which reduces to the latter in 
the perturbative limit. The vertices are obtained as follows.  
Using the above assumption about weak breaking of diffeomorphims invariance we 
first expand the classical action in analogous fashion to 
\eqref{vertex_exp}, which leads to
\begin{equation}
S[\bar{g},h] =\sum_{n=0}^{N} \frac{1}{n!} \left. 
\frac{\delta^n S[\bar{g},h]}{\delta h^n} \right|_{h=0} h^n \,,
\end{equation}
with $S$ given by \eqref{gen_action}.
Again, the notation is symbolic and indices and its contractions as well 
as spacetime integrals are suppressed. Introducing an explicit notation, the 
quadratic part takes the form
\begin{align}
\left. \frac{\delta^2 S[\bar{g},h]}{\delta h^2} \right|_{h=0} 
h^2 = \int_{x_1,x_2} \left(S^{(2)} \right)^{A_1 A_2} 
h_{A_1} h_{A_2} \,,
\end{align}
with the super-index $A_i = \{\mu_i \nu_i,x_i\}$ and the classical two-point 
function
\begin{align}
\left(S^{(2)} \right)^{A_1 A_2} = \left. \frac{\delta^2 
S[\bar{g},h]}{\delta h_{\mu_1 \nu_1}(x_1) \delta h_{\mu_2 \nu_2}(x_2)} 
\right|_{h=0} \,, 
\end{align}
where we have explicitly written out the single components of the super-index 
$A_i$.
\\We proceed by choosing a linear split of the metric
\begin{equation}
g_{\mu \nu}= \bar{g}_{\mu \nu} + \sqrt{G} h_{\mu \nu} \,,
\end{equation}
with $\bar{g}_{\mu \nu} = \mathbb{1}$. Moreover,  
the graviton field $h$ acquires the usual mass dimension one for a bosonic 
field and a canonical kinetic term due to the factor of $\sqrt{G}$ in the 
definition. The classical vertex functions in the 
above expansion acquire a scale dependence due to quantum fluctuations, which 
turns the the gravitational coupling $G$ into a scale-dependent running 
coupling $G_k$. This also leads to a dressing of the 
vertex functions $\Gamma^{(n)}$ with an overall coupling $G_k^{\frac{n}{2}-1}$.
Moreover, we account for 
quantum contributions with a scale-dependent wave-function renormalization 
$Z_k$ for the 
graviton field. From general reparameterization invariance of the effective 
action, it 
follows that if the field scales with $Z_k^{1/2}$, then the vertex 
functions $\Gamma^{n}$ scale as $Z_k^{n/2}$. As a consequence, the vertex 
function $\Gamma^{(2)}$ has an overall renormalization factor $Z_k$, which 
determines the anomalous scaling. In summary, this procedure can be 
reformulated as a rescaling of the classical graviton field according to
\begin{equation}
h \, \, \longrightarrow \, \, \left( Z_k G_k\right)^{\frac{1}{2}} h \,.  
\end{equation}       
So far, this procedure fixes the overall scale-dependence of vertex 
functions. In addition to that, we have to take into account the relative 
scale-dependence of the couplings associated to different tensor structures, 
i.e.\ in our case the couplings of the different diffeomorphism invariants. 
Therefore, we also allow for a scale-dependence of the 
couplings associated with the four-derivative interactions, which corresponds 
to $a \longrightarrow a_k$ and $b \longrightarrow b_k$, as well as a running of 
the cosmological constant $\Lambda_k$. In addition to this, it is important to 
note that due to the non-diffeomorphism invariance and background dependence of 
the vertex functions $\Gamma^{(n)}$, all of the above dressings with a scale 
dependence are in principle different for each order $n$, i.e.\ there is an 
overall gravitational coupling $G^{n}_k$ related to the $n$-graviton vertex and 
similarly running couplings $a^{(n)}_k$, $b^{(n)}_k$ and $\Lambda^{(n)}_k$.  
This reflects the fact that due to the lack of gauge invariance, the couplings 
that belong to different orders of the 
correlation functions are not protected in the sense that gauge invariance 
enforces simple relations amongst them. This is exactly the same 
problem as in Yang-Mills theory, where the perturbative beta-function extracted 
from the one-loop effective action is the same as the one extracted from 
one-loop approximations of the two-, three, or four-point function. However, in 
the non-perturbative regime this not true and different couplings that agree 
in perturbation theory must be distinguished. In the present approximation we 
set $G^{(n)}_k 
\equiv G_k$ for $n\geq3$ and for the constant terms $\Lambda^{(n)}_k \equiv 0$ 
for $n\geq3$. The constant term of the two-point function is 
conveniently written as $M^2_k = - 2 \Lambda^{(2)}_k$ and is called the 
effective graviton mass parameter.  
We demonstrate this construction explicitly for the transverse traceless 
component of the graviton two-point function, which is now an effective, 
dressed 
correlator and is transformed in momentum space according to 
\begin{align} \notag
S^{(2)}_\mathrm{TT}(p_1,p_2) = \frac{1}{32 \pi}(-2 \frac{\Lambda}{G} & + 
\frac{p^2}{G} + b \, 
p^4) 
\delta(p_1+p_2) 
\\ \notag & \downarrow
\\ \Gamma^{(2)}_\mathrm{TT}(p_1,p_2) = \frac{Z_k}{32 \pi} (M_k^2 & + p^2 + 
G_k \, b_k p^4 ) 
\delta(p_1+p_2) \,,
\end{align}
where first, all couplings in $S^{(2)}$ are dressed with a $k$-dependence, a 
wave-function renormalization factor of $Z_k$ is attached according to the 
above general renormalization group arguments, and the entire expression is 
multiplied by $G_k$. Similarly, the three-point function is then obtained from 
the classical vertex $S^{(3)}_{A_1 A_2 A_3}(\Lambda,a,b;p_1,p_2,p_3)$ as
\begin{align}\notag
& \Gamma^{(3)}_{k;A_1 A_2 A_3}(Z_k,G_k,a_k,b_k;p_1,p_2,p_3) =  
\\  & Z^{\frac{3}{2}}_k 
G_k^{\frac{1}{2}} S^{(3)}_{A_1 A_2 A_3}(1,0,G_ka_k,G_kb_k;p_1,p_2,p_3)\,.
\end{align}
In general a vertex function of order $n\geq3$ reads 
\begin{align}\label{vertex}\notag
 & \Gamma^{(n)}_{k;A_1 ... A_n}(Z_k,G_k,a_k,b_k;(\mathbf{p})) = 
\\ & Z_k^{\frac{n}{2}} G_k^{\frac{n}{2}-1} S^{(n)}_{A_1 ... 
A_n}(1,0,G_ka_k,G_kb_k;(\mathbf{p})) \,,
\end{align}
where $(\mathbf{p})=(p_1,...,p_n)$.
\\Crucially, all 
the coupling constants above are fluctuation field couplings, as they are 
related to functional derivatives with respect to the fluctuation field.
Although we identify $G_k^{(n)} \equiv G_k$ for $n\geq3$, we 
resolve the important difference between the graviton wave-function 
renormalization $Z_k$ and Newtons coupling $G_k$ as well as the difference 
between the mass parameter $M^2_k$ of the 
fluctuation field propagator and the background cosmological constant, which is 
given by $\Lambda^{(0)}_k$.  
Finally we mention, that while the fluctuation 
field couplings constitute the dynamical set of parameters of the theory, they 
are not directly related to observables. It is also for that reason, that 
$M^2_k \neq0$ is not to be confused with a model of massive gravity.   
\\In such an expansion around a flat background, the flow equation for the 
propagator is given by a momentum integral over propagators and vertices 
according to
\begin{align}\label{two_point_flow_non_diag}
\notag &\mathrm{Flow}^{(2h)}_{\alpha\beta\mu\nu} = 
\\ & \notag
-\frac{1}{2} \int_{\mathbb{R}^4}
\frac{\mbox{d}^4q}{(2\pi)^4}
\Gamma^{(4h)}_{\alpha\beta\gamma\tau\iota\kappa\mu\nu}(p,q,
-q,-p) \hspace{3 pt} (\mathcal{G}\dot{R}\mathcal{G})_{hh}^{\iota \kappa \gamma 
\tau}(q)
\\  \notag + & \int_{\mathbb{R}^4}
\frac{\mbox{d}^4q}{(2\pi)^4}
\Gamma^{(3h)}_{\alpha\beta\gamma\tau\iota\kappa}(p,q,-p-q) \hspace{2 pt}
(\mathcal{G}\dot{R}\mathcal{G})_{hh}^{\epsilon\delta\gamma\tau}(q) \hspace{2 pt}
\\ \notag & \times
\Gamma^{(3h)}_{\mu\nu\epsilon\delta\rho\sigma}(-p,-q,p+q)
 \hspace{2 pt} \mathcal{G}_{hh}^{\iota\kappa\rho\sigma}(p+q) 
\\ \notag & -2 \int_{\mathbb{R}^4} \frac{\mbox{d}^4q}{(2\pi)^4}
\Gamma^{h \bar{c} c}_{\alpha \beta \gamma \tau}(p,q,-p-q)  
(\mathcal{G}\dot{R}\mathcal{G})_{\bar{c}c}^{\gamma \delta}(q) 
\\ & \times \Gamma^{h \bar{c} c}_{\mu \nu
\rho \sigma}(-p,-q,p+q) \mathcal{G}_{\bar{c}c}^{\tau \rho}(p+q) \,,
\end{align}
where $\mathcal{G}:=(\Gamma^{(2)}_k +R_k)^{-1}$ is the regularized, full 
propagator.
Equation \eqref{two_point_flow_non_diag} is just the explicit form of the 
diagrammatical representation depicted 
in \autoref{fig:flow_of_prop}.        
\\We end this section with some remarks on the running couplings involved in 
the flow equations within the present approximation. On the left hand side of 
\eqref{two_point_flow_non_diag} there appears the scale derivative of the 
two-point function, which in turn contains scale derivatives of the wave 
function 
renormalization $Z_k$, the mass parameter $M_k^2$ and the four 
derivative couplings $a_k$ and $b_k$. This means that the set 
\begin{equation}
\left(\dot{Z}_k, \dot{M}^2_k,\dot{a}_k,\dot{b}_k\right) \, ,
\end{equation}
will define the beta-function of the 
theory. The gravitational coupling $G_k$ will also enter the LHS of equation 
\eqref{two_point_flow_non_diag} for the two-point function, but only in 
combination 
with the four-derivative couplings $a_k$ and $b_k$.
Moreover, it is important to note that the wave-function renormalization $Z_k$ 
is not an essential coupling and its dependence will appear only 
through the anomalous dimension $\eta_k:= -\dot{Z}_k/Z_k$, which will obey an 
algebraic rather than a differential equation. However, the RHS in 
\autoref{fig:flow_of_prop} contains the three- and the four-point vertex, which 
are proportional to $G_k$, whose scale derivative is not determined by the 
equation for the two-point function but by the higher order vertex functions. 
This reflects the fact that in the infinite hierarchy \eqref{flow_hierarchy} 
the flow equation of order $n$ depends on vertex functions up to order $n+2$. 
Therefore, we can treat the coupling $G_k$ as a free parameter in the 
beta-functions and study the dependence parametrically, or we can use equations 
obtained from the three-point function with Einstein-Hilbert tensor structures, 
which amounts to neglecting the feedback of the higher derivative couplings. 
Both possibilities will be taken into account.

\subsection{Gauge fixing}
The standard way of gauge fixing in four-derivative quantum gravity is by 
choosing a gauge fixing condition that is also fourth order in derivatives, 
see e.g. \cite{Codello:2006in,Benedetti:2009rx,Saueressig:2011vn}. In this 
work, we present a different gauge fixing condition, which is second order in 
derivatives. This is sufficient in order to define an invertible two-point 
function and therefore a gauge-fixed propagator. We use the two-parameter 
family of gauge fixing conditions given by    
\begin{equation}
F_\mu = \bar{\nabla}^\nu h_{\mu \nu} -\frac{1+\beta}{4} \bar{\nabla}_\mu 
h^\nu_\nu \,,
\end{equation}
which results via Faddeev-Popov quantization in the gauge-fixing action
\begin{equation}
S_\mathrm{GF} = \frac{1}{2 \alpha} \int \mathrm{d}^4x \sqrt{\text{det}\bar{g}} 
\bar{g}^{\mu 
\nu} F_\mu F_\nu \,.  
\end{equation}
The Landau limit $\alpha \longrightarrow 0$ corresponds to a sharp 
implementation of the gauge fixing condition and is a fixed point of a 
scale-dependent gauge-fixing parameter $\alpha_k$. Moreover, as we will see 
below, the choice $\beta=-1$ diagonalizes the propagator-matrix in the 
Landau-gauge. It has also been argued recently that the choice 
$\alpha \longrightarrow 0$ and $\beta \longrightarrow -1$ corresponds to the 
``physical gauge-fixing'' as it acts on true gauge fluctuations only,  
\cite{Wetterich:2016vxu,Wetterich:2016ewc,Wetterich:2016qee}. The diagonal 
structure of the propagator in this gauge will be made explicit in the next 
section. 
\\Moreover, exponentiation of the Fadeev-Popov determinant introduces the 
Grassmann-valued ghost fields $\bar{c}$ and $c$. Their action is given by
\begin{equation}
\int\text{d}^4x \sqrt{\text{det}\bar{g}}\; \bar{c}^\mu
\mathcal{M}_{\mu\nu} c^\nu \,,  
\end{equation}
with the Fadeev-Popov operator
\begin{equation}
M^{\mu}_{\hphantom{\mu}\nu} =
\bar{g}^{\mu \alpha}\bar{\nabla}^{\beta} \left(g_{\alpha \nu}
  \nabla_{\beta} + g_{\beta \nu} \nabla_{\alpha} \right) - \frac{1}{2}(1+\beta)
\bar{g}^{\alpha \beta} \bar{\nabla}^{\mu} g_{\beta \nu}
\nabla_{\alpha} \,.
\end{equation}

\subsection{Two-Point Function, Regulator and Propagator}

With the vertex construction introduced in the previous 
section and the full basis of four-derivative operators we can now derive the 
components of the graviton-two-point function. In what follows, we drop the 
subscript $k$ for the scale-dependent couplings for better readability. Unless 
stated otherwise, all couplings and vertex functions are from now on 
scale-dependent. We expand the two-point function in a complete set of 
projectors according to   
\begin{equation}\label{Gamma_proj}
\Gamma^{(2)} = \sum_{i=1}^6 \Gamma^{(2)}_i \, P_i \,.
\end{equation}
The pseudo-projectors $P_i$ are introduced in the appendix, 
\autoref{app_inversion}. For 
general gauge fixing parameters the different components of the graviton 
two-point functions are then given by
  
 \begin{equation}\label{Gamma2TT}
\Gamma^{(2)}_{1}=\Gamma^{(2)}_{\mathrm{TT}} = \frac{Z}{32 \pi}(M^2 + 
p^2 + G \,b \, 
p^4) \, ,     
\end{equation}

\begin{equation}\label{Gamma2V}
\Gamma^{(2)}_{2}=\Gamma^{(2)}_{\mathrm{V}} = \frac{Z}{32 \pi}\left(M^2 + 
\frac{p^2}{\alpha} 
\right) \, ,    
\end{equation}

\begin{align}\label{Gamma2Tr}
\Gamma^{(2)}_{3} = \frac{Z}{32 \pi}\frac{1}{8 
\alpha} \Big( & -4 \alpha M^2 + p^2( -16 \alpha +3 (\beta+1)^2)  
\\ \nonumber  & + 32 p^4 \alpha 
(3 a + b) \Big) 
\end{align}

\begin{equation}\label{Gamma2WW}
\Gamma^{(2)}_{\mathrm{4}} = \frac{Z}{32 \pi}\frac{1}{8 
\alpha} \Big(4 \alpha M^2 + p^2(\beta-3)^2 \Big) \, ,    
\end{equation}

\begin{equation}\label{Gamma2WS}
\Gamma^{(2)}_{\mathrm{5}} = \Gamma^{(2)}_{\mathrm{6}} = \frac{Z}{32 
\pi}\frac{\sqrt{3}}{8 
\alpha} \Big(- 4 \alpha M^2 + p^2(\beta-3) (\beta+1) \Big) \, .    
\end{equation}
As it is well-known, the contributions of the $R^2$ operator to the $TT$ 
component vanish, and therefore the $p^4$ coefficient originates only from the 
$R_{\mu \nu}^2$, or $C^2$, term.

In the physical gauge the inverse two-point function takes the form
\begin{equation}
\left( \Gamma^{(2)} \right)^{-1} = \left( \Gamma^{(2)} 
\right)_1^{-1} P_1 + \left( \Gamma^{(2)} 
\right)_3^{-1} P_3 \,,  
\end{equation}
i.e.\ all but the $TT$- and one scalar component of the graviton propagator 
vanish. Explicitly, these components read
\begin{equation}
\left( \Gamma^{(2)} \right)_1^{-1} = 32 \pi Z \frac{1}{M^2 + p^2 + G \, b 
p^4} \, ,
\end{equation}

\begin{equation}
\left( \Gamma^{(2)} \right)_3^{-1} = 32 \pi Z \frac{2}{-M^2 - 4 p^2 + 8 G 
p^4(3 a +b)} \,, 
\end{equation}
where we immediately identify the well-known factor $(3 a +b)$ attributed to 
the conformal combination $R_{\mu \nu} R^{\mu \nu} - 1/3 R^2$.
\\The choice of the regulator is a crucial ingredient in the construction of 
the renormalization group flow. We choose a regulator that enables us 
to do completely analytical calculations. Such a regulator is given by the  
Litim regulator
\begin{equation}\label{regulator}
\left( R_k(q^2) \right)_i = (\Gamma^{(2)}_i(k)-\Gamma^{(2)}_i(q)) 
\theta(k^2-q^2) 
P_i \,, 
\end{equation}
adjusted for each component of the two-point function $\Gamma^{(2)}_i(q^2)$. 
This 
can easily be 
rewritten as
\begin{equation}
\left(R_k(q^2) \right)_i = \Gamma^{(2)}_i r_k(q^2) P_i
\end{equation}
with the dimensionless shape function    
\begin{equation}
r_k(q^2)= \left(\frac{\Gamma_i^{(2)}(k^2)}{\Gamma_i^{(2}(q^2)}-1\right) 
\theta(k^2-q^2) 
\,.  
\end{equation}
It is worth noting, that the Landau limit $\alpha \longrightarrow 0$ cannot be 
taken in the regulator $R$, but the propagator $\mathcal{G}$ and the product of 
propagators with the scale-derivative regulator $\mathcal{G}\dot{R}\mathcal{G}$ 
that enter the flow equations are finite in this limit.

\section{Beta-Functions}

With the construction of the vertices, the propagator and the regularization of 
the last sections, we are now in a position to derive the $\beta$-functions
$\left(\beta_{M^2},\beta_a,\beta_b\right)$ and the anomalous dimension $\eta$
via a suitable projection of the flow equation 
\eqref{two_point_flow_non_diag}.

\subsection{Projection}
From the equations 
(\ref{Gamma2TT},~\ref{Gamma2V},~\ref{Gamma2Tr},~\ref{Gamma2WW},~\ref{Gamma2WS}) 
for the two-point functions one can see that the TT-mode is independent of the 
gauge fixing, as is the $p^4$ coefficient of the trace-mode. Moreover, as 
we have already pointed out, the $p^4$ coefficient of the former receives 
contributions from the $\mathrm{Ric}^2$ operator only, while the latter also 
contains contributions from the $R^2$ term. We exploit this fact for the 
definition of the 
projections on the running couplings. Introducing the operator $\circ$ 
that denotes full contraction of tensor indices, we obtain
\begin{equation}\label{beta_mu}
\partial_t \mu = -2 \mu + \mu \, \eta + \frac{32 \pi}{5k^2} \lim_{p 
\rightarrow 0} \left( \mathcal{P}_{\mathrm{TT}} \circ 
\mathrm{Flow}^{(2)} \right) \, ,
\end{equation}
for the running of the mass parameter,
\begin{equation}\label{beta_eta}
\eta = - \frac{16 \pi}{5} \lim_{p 
\rightarrow 0} \frac{\partial^2}{\partial p^2} \left( \mathcal{P}_{\mathrm{TT}} 
\circ
\mathrm{Flow}^{(2)} \right) \, ,   
\end{equation}
for the anomalous dimension $\eta$ and
\begin{align}\label{beta_a}
\partial_t a = & -\frac{\partial_t b}{3} + 
\frac{(3\,a+b)(g(2+\eta)-\partial_t g)}{3\,g} 
\\ \nonumber & + \frac{\pi k^2}{9 \,g} \lim_{p 
\rightarrow 0} \frac{\partial^4}{\partial p^4} \left( \mathcal{P}_{\mathrm{Tr}} 
\circ \mathrm{Flow}^{(2)} \right) \,,  
\end{align}
as well as
\begin{align}\label{beta_b}
\partial_t b = \frac{b}{g} (g(2 + \eta)- \partial_t g) + \frac{4 \pi k^2}{15 \, 
g} \lim_{p 
\rightarrow 0} \frac{\partial^4}{\partial p^4} \left( \mathcal{P}_{\mathrm{TT}} 
\circ \mathrm{Flow}^{(2)} \right) \, ,     
\end{align}
for the four-derivative couplings. 
\\There are some subtleties concerning the above momentum projection. As 
described in the previous section we employ a Litim cutoff. It is 
well-known that this cutoff does not allow an expansion in powers of $p^2$ of 
the right-hand side of the flow equation, but with such a cutoff the flow 
is an expansion in the absolute value $p$, where odd powers of $p$ appear 
beyond quadratic order \cite{Morris:2005ck}. 
However, we expect that this is not a problem here, as we can always introduce 
a smooth version of the Litim-cutoff by the replacement $\theta_\epsilon$ 
with $\theta_\epsilon \longrightarrow \theta$ in the limit $\epsilon 
\longrightarrow 0$, i.e.\ just a smeared version of the $\theta$-function. For 
any finite $\epsilon$ the flow does allow an expansion in $p^2$ and 
the limit $\epsilon \longrightarrow 0$ exists on both sides of the flow 
equation. Therefore, we expect that there should be no qualitative difference 
between the Litim regulator and its smoothened counterpart.  
Nevertheless, it is 
certainly true that the cutoff employed in this work is not optimized in the 
present fourth-order approximation. Optimization and the convergence of the 
derivative expansion is discussed 
in \cite{Litim:2000ci,Pawlowski:2005xe,Morris:1999ba}.    
\\There are further technical difficulties arising due to derivatives of 
the $\theta$-function. The momentum derivatives of the $\theta$-function 
immediately produce $\delta$-functions and derivatives thereof. In the 
limit $p \longrightarrow 0$ these distributional products are not well-defined. 
However, with a proper treatment these ill-defined terms do not appear. These 
mathematical problems are solved in the appendix, \autoref{app_mom_proj}.

\subsection{Fixed Points of Non-Perturbative Beta-functions}

The construction described in the last sections leads to the set of $\beta$ 
functions for the couplings $a$, $b$ and $M^2$, as well as an algebraic 
equation for the anomalous dimension. These equations also depend on the 
gravitational constant $G$, which is in principle obtained from the 
three--point function $\Gamma^{(3)}$. First, we close this system of equations  
by using an equation for the beta-function $\beta_G$ of the gravitational 
coupling obtained from the flow equation for $\Gamma^{(3)}$ in a vertex 
expansion with Einstein-Hilbert tensor structures, \cite{Christiansen:2015rva}.
With this $\beta$-function for 
the gravitational coupling we ignore the direct feedback of the higher 
derivative couplings $a$ and $b$ to the running of the gravitational coupling. 
More precisely, in a fully consistent calculation where the gravitational 
coupling is obtained from a three--point function including higher derivative 
structures, there will be terms proportional to $a$ and $b$ in the 
$\beta$-function for g. Here, the feedback is only indirect via the dependence 
of $\mu$ and $\eta$ on $a$ and $b$.    
The fixed point analysis is then formulated for the dimensionless 
couplings, and we define $g=Gk^2$ and $\mu = M^2/k^2$. 
In total, we obtain the set of equations
\begin{equation}
\left(\dot{g},\dot{Z}, \dot{\mu},\dot{a},\dot{b}\right) \, .
\end{equation}
These $\beta$-functions are all derived in closed analytic form, however the 
expressions are way too bulky in order to be given explicitly in this paper.   

The fixed point condition for this set of couplings  
reads $\beta_a = \beta_b= \beta_g = \beta_\mu =0$, whereas the 
anomalous dimension $\eta$ takes a value dynamically determined by the fixed 
point values of the couplings. The number of relevant and irrelevant 
directions is determined by the properties of the linearized flow around the 
fixed point, which, in turn is characterized by the eigenvalues $\theta_i$ of 
the stability matrix $B$. The stability matrix is given by
\begin{equation}
B_{ij} = \left. \frac{\partial \beta_{g_i}}{\partial g_j} 
\right|_{g_i=g_i^*}\,, 
\end{equation}
where $\{g_i\}$ represents the set of all coupling constants.
Negative eigenvalues $\theta_i$ of the stability matrix indicate a relevant 
direction, while positive eigenvalues belong to a irrelevant direction. An 
irrelevant direction implies that one parameter is fixed by the asymptotic 
safety condition. More precisely, one initial condition of the flow is 
fixed and therefore the evolution of the coupling with energy is determined by 
the theory. Consequently, the value of this coupling constant at an arbitrary 
energy scale is a prediction of the theory. 

An analytical solution of the full equations is not 
possible, but numerically the system exhibits several fixed points. However, 
only one fixed point has eigenvalues 
$\theta_i<10$ of the stability matrix and obeys the constraints $g_*>0$ and 
that all couplings are real valued.  
This fixed point has the coordinates   
\begin{align}\label{Non-Pert_FP}
\left(g^*, \mu^*, a^*, b^*, \eta^*  \right) = \left(0.43, -0.34 , -0.41, 0.91, 
0.77  \right) \,.
\end{align}
with eigenvalues of the stability matrix given by
\begin{align}\label{Non-Pert_FP_Crit_Exp}
\left(\theta_1, \theta_2, \theta_3, \theta_4 \right) = \left(-1.5 - 2.7 
\, i, -1.5 + 2.7 \, i, 2.4, 8.3 \right) \,.
\end{align}
A very important property of this fixed point is that it is characterized by 
two irrelevant and two relevant directions. In all Einstein-Hilbert like 
approximations, i.e.\ where one retains only two couplings, one finds two 
negative eigenvalues, i.e.\ two relevant directions. In our case, we included 
the four-derivative couplings, but the UV-critical surface remains 
two-dimensional. Interestingly, our result differs in this respect from the 
structure of the non-Gaussian fixed point in four-derivative gravity found 
within background field flows, where the UV-critical surface is 
three-dimensional, \cite{Benedetti:2009rx}. However, we note that including the 
full feedback of the higher derivative couplings into the $\beta$-function for 
the gravitational coupling can in principle turn an irrelevant direction into a 
relevant one.
In the light of the asymptotic safety conjecture, our  
result is very encouraging, as predictivity is encoded in a finite dimensional 
UV-critical surface. The four-derivative couplings $a$ and $b$ are classically 
marginal, but quantum corrections turn them into irrelevant couplings. 
Couplings related 
to even higher derivatives are classically irrelevant, and increasingly large 
quantum fluctuations would be necessary in order to form further relevant 
directions. Based on our results, it is reasonable that this does not happen in 
the present case, as 
even the marginal couplings turn irrelevant.  
Moreover, high order polynomial expansions of $f(R)$ truncations 
with background field flows show near-Gaussian scaling for higher order 
operators \cite{Falls:2013bv,Falls:2014tra}, and it is reasonable that this 
pattern translates also to the fluctuation field flow equations.

\subsection{Stability Check: Expansion of Threshold Functions and Corresponding 
Fixed Points}

In the non-perturbative beta-functions above, resummations to infinite order 
are included by the anomalous dimensions on the right-hand side of the flow 
equations, as well as by the non-trivial $g$-dependence in the propagators. 
Both 
aspects turn the right-hand side into a infinite power series in $g$, whose 
convergence depends on the value of the coupling. 
There are two aspects one needs to take into account. First of all, the 
complicated structure of the resummations obviously induces many fixed 
points, which might either have unphysical properties or are truncation 
artefacts. This is particularly important as the resummations are all regulator 
dependent.
The potentially dangerous ones are the latter ones, as they appear 
at first sight as fixed points with viable physical properties, such as 
positive 
Newton coupling and real-fixed point values of all couplings. Nevertheless, 
some of 
them might be only present in the truncated theory and will disappear once 
further improvements of the truncation are taken into account. Sometimes such 
fixed points reveal themselves by very large critical exponents. However, 
it has also been observed, that by including higher order operators and new 
couplings, the critical exponents can grow quite large, but converge to 
smaller values once even further improvements of the truncation are taken into 
account \cite{Falls:2013bv,Falls:2014tra}. This makes it difficult to ensure 
that a fixed point does not 
fall into this class. We rush to mention that the fixed point found in the 
previous section \eqref{Non-Pert_FP} with the moderate eigenvalues 
\eqref{Non-Pert_FP_Crit_Exp} is not in this class, nevertheless, it is 
important to check the reliability of this fixed point. 
One way of doing this is to expand the right hand-sides of the 
$\beta$-functions in powers of the resummation parameter, i.e.\ in powers of 
the coupling $g$ in our case. This removes artificial zeroes of the beta 
functions, however, in principle it can also remove fixed points that are 
induced by non-perturbative effects. In addition to that, this procedure 
makes sense only by expanding to lowest order, as at higher orders artificial 
zeros are created again, and for obvious reasons polynomial expansions are 
particularly dangerous in that respect. 
\\Therefore, we expand the full beta functions 
(\ref{beta_mu},~\ref{beta_eta},~\ref{beta_a},~\ref{beta_b}) 
 to first order in $g$ and solve the fixed point equations. As a result we find 
\begin{align} \label{FP_without_resum}
\left(g^*, \mu^*, a^*, b^*, \eta^*  \right) = \left(0.59, -0.29 , -0.35, 0.51, 
0.56  \right) \,,
\end{align}
with critical exponents
\begin{align}
\left(\theta_1, \theta_2, \theta_3, \theta_4 \right) = \left(-2.2 + 2.3 
\, i, -2.2 - 2.3 \, i, 2, 2.5\right) \,,
\end{align}
which agrees rather well with the fixed point \eqref{Non-Pert_FP}. 
Interestingly, all other fixed points have at least one coupling constant with 
a complex fixed point value and are therefore unphysical. This result provides 
evidence that the 
fixed point of the full equations including all resummations is not just a 
truncation artefact.

\subsection{Parametric Study of the Fixed Point}

In the fixed point analysis presented above we have used an equation for 
$\beta_g$ from the flow of the three-point function, but with Einstein-Hilbert 
tensor structures. We have already mentioned that this is an approximation 
where there is no direct feedback of the four-derivative 
couplings to $\beta_g$, but only an indirect feedback via $\eta$ and 
$\mu$. Therefore, it 
is interesting to study the behavior of the fixed point by treating $g$ as a 
free parameter. This means that we solve the system of equations given by 
$\beta_a = \beta_b= \beta_\mu =0$ and the algebraic equation for the anomalous 
dimension $\eta$. We find a continuous deformation of the fixed point 
\eqref{FP_without_resum} in the range $g \in [0,\approx 0.7]$, where in the 
limit $g \longrightarrow 0$ the fixed point turns into the Gaussian fixed 
point, whereas for $g \gtrapprox 0.7$ it turns into a pair of complex conjugate 
fixed points.

\subsection{One-Loop Beta-Functions and Perturbative Fixed Point}

As we have already discussed, a classical theory of gravity based on the higher 
derivative action \eqref{gen_action} is perturbatively renormalizable due to 
the $p^{-4}$-propagator. Moreover, the coupling constants $a$ and $b$ are 
dimensionless, and therefore universal, in the sense that they 
are independent of the regularization. The perturbative 1-loop running of the 
$\mathrm{Ricc}^2$ coupling $b$ is given by the beta-function
\begin{equation}\label{pert_b}
\beta^{\mathrm{pert}}_b =  \frac{1}{(4 \pi)^2} \frac{133}{10} \approx 0.084 \,.
\end{equation}
As this beta function is constant there is no fixed point on the perturbative 
one-loop level in the coupling $b$. However, by changing the basis in the space 
of four derivative operators to the parameterization 
\eqref{gen_action_reparam}, one finds in four dimensions $b=s^{-1}$, and 
therefore
\begin{equation}
\beta^{\mathrm{pert}}_s = - \frac{1}{(4 \pi)^2} \frac{133}{10} s^2 \,, 
\end{equation}
i.e.\ the coefficient is negative and therefore the coupling is asymptotically 
free. This result is found within perturbative calculations for the one-loop 
effective action \cite{Fradkin:1981hx,Avramidi:1985ki,Niedermaier:2009zz} and 
with the Functional 
Renormalization Group and background field flows 
\cite{Benedetti:2009rx,Codello:2006in}.
In the present setup, the calculation differs from these, as we calculate the 
fluctuation field couplings. In our setup the one-loop coefficient reads
\begin{equation}\label{pert_b_fluc}
\beta^{\mathrm{pert}}_b = \frac{29}{32 \pi^2} \approx 0.092 \,, 
\end{equation}
which differs from \eqref{pert_b} by a factor of $1.1$.    
Therefore, the one-loop coefficient is not universal in the sense that 
background and fluctuation field calculations agree. The reason for this 
disagreement is most likely rooted in the non-triviality of Nielsen identities 
at this fixed point. Also gauge- and parameterization dependencies might play a 
role here. This very important issue is under current investigation and the 
results will be reported elsewhere.

\section{Conclusions}

In this work we have studied the graviton propagator of four-derivative quantum 
gravity non-perturbatively. We used an expansion of the effective action in 
terms of the dynamical graviton field and flow equations for the vertex 
functions derived from the Wetterich equation. This enabled us to determine the 
scale dependence of the Einstein-Hilbert and four-derivative couplings as well 
as the graviton anomalous 
dimension. We analyzed the resulting $\beta$-functions and found a non-Gaussian 
fixed point, which can be understood as an extension of the well-known fixed 
point in Einstein-Hilbert gravity. An important property of this fixed point is 
that there are only two relevant directions, i.e.\ the same number as in the 
Einstein-Hilbert truncation. This means that the two classically marginal 
operators $R^2$ and $R_{\mu \nu}^2$ do not generate further relevant 
directions. This is of particular interest as a relevant direction 
always corresponds to a free parameter, which needs to be fixed by 
external input, e.g.\ some measurement. Therefore, a fixed point describes a  
predictive theory only if the number of relevant directions is finite. 

Based on this work, there are several important directions that one can 
pursue. First of all, it would be interesting to include even further 
invariants in order to test the conjecture that classically irrelevant 
operators stay irrelevant. The present work provides the first step in 
systematically including more tensor structures in vertex expansions. 
Moreover, a genuine flow of the gravitational 
coupling $g$ from the graviton three-point and four-point function along the 
lines presented in 
\cite{Christiansen:2015rva} but with higher derivative operators would provide 
very valuable insights. These two studies are of major importance in order to 
test an apparent convergence in vertex expansions, see also \cite{4_pt}. 
Furthermore, the stability of derivative 
expansion around $p=0$ should be tested in a systematic way. 
 In addition to that, the flow away 
from the fixed point towards the infrared encodes the information whether the 
non-trivial ultraviolet fixed point is indeed connected to infrared physics 
that describes general relativity. This is essential for the 
question if the non-Gaussian fixed point is a viable candidate for the 
construction of the continuum limit. In order to connect with reality, it is 
also inevitable to study the coupling to gauge and matter fields.    
\\[-1.5ex]       

\noindent {\bf Acknowledgements} The author wants to thank A.~Eichhorn, 
K.~Falls, J.M.~Pawlowski, M.~Reichert, F.~Saueressig and C.~Wetterich for many 
valuable 
discussions. NC acknowledges funding by the DFG, Grant EI 1037-1.


\appendix
\allowdisplaybreaks

\section{Tensor decomposition of the graviton propagator}\label{app_inversion}
As the two-point function needs to be 
inverted in order to obtain the propagator, we represent $\Gamma^{(2)}$ in a 
quasi-projector basis given by the six projection operators $P_{\mathrm{TT}} = 
P_1$, $P_{\mathrm{V}} = 
P_2$, $P_3$, $P_4$ and $P_5$, $P_6$.
The index $TT$ indicates that this operator projects on the 
transverse-traceless tensor structure of a symmetric rank two tensor, and 
analogously $V$ refers to the vector mode, while $P_3$ and $P_4$ project on two 
scalar components. The operators $P_5$ and $P_6$ 
generate mixings in the scalar sector, as we will see below.
In terms of the well-known transverse and longitudinal projectors
\begin{equation}
\Pi_{T,\mu \nu}(p) = \delta_{\mu\nu} - \frac{p_{\mu}p_{\nu}}{p^2} \,, 
\end{equation}
and
\begin{equation}
\Pi_{L,\mu \nu}(p) = \frac{p_{\mu}p_{\nu}}{p^2} \, 
\end{equation}
the above operators read
\begin{align}\nonumber
P_{1,\mu \nu \alpha\beta}(p) = & \frac{1}{2}\left(\Pi_{T,\mu \alpha}(p) 
\Pi_{T,\nu \beta}(p) + \Pi_{T,\mu \beta}(p)\Pi_{T,\nu \alpha}(p) \right)
\\ & -\frac{1}{3} 
\Pi_{T,\mu \nu}(p) \Pi_{T,\alpha \beta}(p)\,,  
\end{align}
\begin{align}\nonumber
P_{2,\mu \nu \alpha\beta}(p) = & \frac{1}{2} ( \Pi_{T,\mu \alpha}(p) 
\Pi_{L,\nu \beta}(p) + \Pi_{T,\mu \beta}(p)\Pi_{L,\nu \alpha}(p)
\\ &  +\Pi_{T,\nu \alpha}(p) \Pi_{L,\mu \beta}(p) + \Pi_{T,\nu \beta}(p) 
\Pi_{L,\mu \alpha}(p))\,,\end{align}
\begin{align}
P_{3,\mu \nu \alpha\beta}(p) = \frac{1}{3} 
\Pi_{T,\mu \nu}(p) \Pi_{T,\alpha \beta}(p)\,,  
\end{align}
\begin{align}
P_{4,\mu \nu \alpha\beta}(p) =  
\Pi_{L,\mu \nu}(p) \Pi_{L,\alpha \beta}(p)\,,  
\end{align}
\begin{align}
P_{5,\mu \nu \alpha\beta}(p) =  \frac{1}{\sqrt{3}}
\Pi_{T,\mu \nu}(p) \Pi_{L,\alpha \beta}(p)\,,  
\end{align}
\begin{align}
P_{6,\mu \nu \alpha\beta}(p) =  \frac{1}{\sqrt{3}}
\Pi_{L,\mu \nu}(p) \Pi_{T,\alpha \beta}(p)\,.  
\end{align}

There is an 
orthogonal subset of these projectors given by 
\begin{equation}\label{op_algebra}
P_i P_j = \delta_{ij} P_j \,\,\,\,\, \mathrm{with} \,\,\, i={1,2,3,4} \,. 
\end{equation}
The relations for the transfer operators $P_5$ and $P_6$ are a bit more 
complicated. It is 
advantageous to introduce two more indices and map the old index $i 
\in (3,4,5,6)$ 
to the index set $(a,b) \in \left( (3,3), (3,4), (4,3), (4,4) \right)$ such 
that the operators in the scalar 
sector can be grouped into a two-by-two matrix. The mapping is done such 
that $P_3=P^{33},P_4=P^{44},P_5=P^{34},P_6=P^{43}$. The relations of the scalar 
projectors in this language read  
\begin{equation}
P^{ab} P^{cd} = \delta^{bc} P^{aa} \,\,\,\, \forall d \,\,\, \mathrm{and} 
\,\,\, a \neq b, c \neq d\,
\end{equation}
\begin{equation}
P^{aa} P^{bc} = \delta^{ab} P^{ac} \,\,\,\, \mathrm{with} \,\,\,b \neq c 
\end{equation}
\begin{equation}
P^{ab} P^{cc} = \delta^{bc} P^{ac} \,\,\,\, \mathrm{with} \,\,\,a \neq b \,,
\end{equation}
and all scalar operators are orthogonal to $P_1$ and $P_2$.
As one can see from these operator relations, the off-diagonal operators 
$P_5=P^{34}$ and $P_6=P^{34}$ induce a 
mixing in the scalar sector and therefore they are also called the spin-zero 
transfer 
operators. As the operators $P_i$ form a complete set, the two point 
function can be expanded as  
\begin{equation}\label{Gamma_proj}
\Gamma^{(2)} = \sum_{i=1}^6 \Gamma^{(2)}_i \, P_i \,.  
\end{equation}
Obviously, this operator set is not orthogonal in the scalar sector due to 
the non-trivial algebra. The coefficients in the above expansion are obtained 
as follows. We define a four-by-four matrix $a$ with diagonal elements 
according to
\begin{equation}
\Gamma^{(2)}_i= a_{ii} = \frac{\mathrm{Tr} \left(P_i \Gamma^{(2)} P_j 
\right)}{\mathrm{Tr} 
\left( P_i P_j \right)} \,\,\,\, \mathrm{with} \,\,\, i \in (1,2,3,4) \, . 
\end{equation}
and off-diagonal elements $a_{34}$ and $a_{43}$. Using the operator relations 
above one can easily obtain that $a_{34} = \Gamma^{(2)}_5 =\mathrm{Tr} P_6 
P_3 \Gamma^{(2)} P_4$ and
$a_{43} = \Gamma^{(2)}_6 =\mathrm{Tr} P_5 P_4 \Gamma^{(2)} P_3$. This 
four-by-four 
matrix can then be inverted and it is easy to show that   
\begin{equation}
\left( \Gamma^{(2)} \right)^{-1} = \sum_{i=1}^6 \left( \Gamma^{(2)} 
\right)_i^{-1} P_i \,,  
\end{equation}
with the coefficients $\left( \Gamma^{(2)} \right)_i^{-1}$ obtained from the 
inverse 
coefficient matrix $a^{-1}$, is indeed the inverse of $\Gamma^{(2)}$.

\onecolumngrid

\section{Momentum Derivatives and Distributions} \label{app_mom_proj}

In this appendix we will derive a general formula for projecting on the $p^2$ 
and the $p^4$ coefficients of the flow at zero momentum and with a
Litim regulator. The derivation is readily generalized for higher order 
derivatives. Moreover, the following derivation can be applied for general 
theories, not 
only for gravity. In order to emphasize the structure of the following 
calculation, we write the propagator in the form
\begin{equation}\label{p4_prop}
G(p) = \frac{1}{Z (M^2 + \alpha p^2 + \beta \, G p^4)(1+r(\frac{p^2}{k^2}))} 
\,, 
\end{equation}
with a mass term $M^2$ and coefficients $\alpha$ and $\beta$. The regulator 
shape function for the Litim regulator and the above propagator 
then takes the form
\begin{equation}\label{shape_func}
r\left(\frac{p^2}{k^2}\right) = \left(\frac{(M^2 + \alpha k^2 + \beta \, G
k^4)}{(M^2 + \alpha p^2 + \beta \, G p^4)} -1\right) \theta(k^2-p^2) \,. 
\end{equation}
All the components of the graviton propagator are of the form \eqref{p4_prop} 
times a tensor structure. Therefore, it is sufficient for the following general 
analysis to assume such a form of the propagator. The main goal is to find a 
closed expression for
\begin{equation}
\lim_{p\rightarrow0}\frac{\partial^n}{\partial p^n} \left( \mathcal{P} 
\circ \mathrm{Flow}^{(2)} \right) \,, 
\end{equation}
where $n=2,4$ and $\mathcal{P}$ is an operator that projects on a 
component of the propagator, which is of the 
form \eqref{p4_prop}. We will also see that the generalization 
to 
arbitrary $n$ is straightforward. 
\\The RHS of equation \eqref{two_point_flow_non_diag} is proportional 
to $\dot{R}$, which is given by
\begin{align}
\dot{R}(q^2) = \left(\dot{\Gamma}^{(2)}(k) - \dot{\Gamma}^{(2)}(q) 
\right) \theta(1-\frac{q^2}{k^2}) 
+ 2 \frac{q^2}{k^2} \left(\Gamma^{(2)}(k) - 
\Gamma^{(2)}(q) \right) \delta(1-\frac{q^2}{k^2}) 
= \left(\dot{\Gamma}^{(2)}(k) 
- \dot{\Gamma}^{(2)}(q) \right) 
\theta(1-\frac{q^2}{k^2}) \,,    
\end{align}
where the last equals sign is of course understood in the distributional sense. 
Therefore, the RHS of the flow equation will always be 
proportional to $\theta(1-\frac{q^2}{k^2})$. Consequently, all theta functions 
$\theta(1-\frac{q^2}{k^2})$ that appear in the propagators 
can be set to one, as the loop integral vanishes for $q^2>k^2$.
This structure makes the tadpol diagram in the flow equation 
\eqref{two_point_flow_non_diag} very easy, as there is no momentum-dependent 
propagator and the momentum derivatives hit only the vertex. 
The derivatives of the contracted tadpol diagram can then be written as
\begin{equation}
\int_{\mathbb{R}^4} \frac{\mbox{d}^4q}{(2\pi)^4}
\Theta(k^2-q^2) \frac{1}{\left(M^2 + \alpha k^2 + \beta G 
k^4 \right)^2} \frac{\partial^n}{\partial p^n} f_1(p,q) \,,
\end{equation}
where $f_1$ depends on the vertex and is a regular function of its arguments. 
Moreover, $f_1$ also depends on the couplings and the anomalous dimensions, but 
this dependence is irrelevant here.
In the self-energy 
diagram, there are momentum dependent propagators $G(p+q)$ and 
corresponding $\theta$ functions. These are treated with a 
case-by-case analysis, i.e., we distinguish the cases wehre the 
$\theta$-function with argument $(p+q)$ in the propagator $G(p+q)$ is either 
one or zero.  
Then, after contraction with a projector, the integrand of the self-energy 
diagram in the flow 
equation \eqref{two_point_flow_non_diag} with a propagator parameterized as 
\eqref{p4_prop} takes the form
\begin{align}
f_2(p,q) \times \begin{cases}\left[ 
\left(M^2 +\alpha k^2 + \beta \, G k^4\right)^3 \right]^{-1} \,\,\,\, & 
\mathrm{if} 
\,\, \theta(k^2 - (p_\mu + q_\mu)^2)=1 
\\
\left[ \left(M^2 +\alpha k^2 + \beta \, G k^4\right)^2 \{ (M^2 + \alpha (p_\mu 
+q_\mu)^2 + 
\beta \, G(p_\mu +q_\mu)^4) \} \right]^{-1} & \mathrm{if} \,\, \theta(k^2 - 
(p_\mu 
+ q_\mu)^2)=0 \,,
\end{cases} 
\end{align}
where $f_2$ is again a function that depends on the vertices and is a regular. 
Due to this regularity, the explicit form of $f_2$ is not relevant 
for the following.
We then introduce the 
definitions  
\begin{equation}
c_1 := \frac{1}{\left(M^2 +\alpha k^2 + \beta \, Gk^4\right)^2} \,,
\end{equation}
and
\begin{equation}
c_2 = \hat{c}_2 \check{c}_2 \hspace{8 pt} \mbox{and} \hspace{10
pt} c_3 = \hat{c}_3 \check{c}_3 \,,
\end{equation}
with
\begin{equation}
\hat{c}_2(q) := \frac{1}{\left(M^2 +\alpha k^2 + \beta \, G k^4\right)^3}
\end{equation}
and
\begin{equation}
\check{c}_2(p,q,x) :=
\Theta\left(k^2-\left(p_\mu + q_\mu\right)^2\right) \,.
\end{equation}
Analogously
\begin{equation}
\hat{c}_3(p,q,x) := \frac{1}{
\left(M^2 +\alpha k^2 + \beta \, Gk^4\right)^2 \left(M^2 + \alpha \left(p_\mu + 
q_\mu\right)^2+\beta \, G\left(p_\mu + 
q_\mu\right)^4 \right)} \,,
\end{equation}
and
\begin{equation}
 \hspace{10 pt} \check{c}_3(p,q,x) := \Theta\left(\left(p_\mu +
q_\mu\right)^2-k^2\right) \hspace{2 pt},
\end{equation}
where $x=\cos(\theta)$, and $\theta$ is the angle between the external 
momentum $p_\mu$ and the loop momentum $q_\mu$, whose absolute values are 
denoted as $p$ and $q$ respectively.
The $p$-dependent terms in the $q$-integrals in the flow, i.e.\,, those which 
depend on the external momentum, are then given by
$f_1,f_2,\check{c}_2,\hat{c}_3,\check{c}_3$.
\\The $n$-th momentum derivative of the flow is then expressed as
\begin{equation}
\lim_{p\rightarrow0}\frac{\partial^n}{\partial p^n} \left( \mathcal{P} 
\circ \mathrm{Flow}^{(2)} \right) \sim 
\lim_{p\rightarrow0} \frac{\partial^n}{\partial p^n} \int_{0}^{\infty} 
\int_{-1}^{1} \mathrm{d}q \, \mathrm{d}x \, q^3 \sqrt{1-x^2}  \theta(k^2-q^2)   
\, 
\left(-\frac{1}{2} f_1 c_1 + f_2 (\hat{c}_2\check{c}_2 + \hat{c}_3 \check{c}_3) 
\right) \,,    
\end{equation}
where we have transformed to 4-d spherical coordinates. First we will 
evaluate this for $n=2$. In order to do so, we 
interchange the $p$-derivatives with 
the $q$-integral at fixed, finite $p$, which is perfectly well-defined. 
Taking two derivatives of the integrand yields 
\begin{align}
\nonumber & \partial^{2}_{p}
\left\{-\frac{1}{2}f_1c_1 +
f_2 \left( \hat{c}_2\check{c}_2 + \hat{c}_3 \check{c}_3 \right) \right\}
\\ \nonumber & = \partial_{p} \left\{ -\frac{1}{2} f_1'c_1 + f_2'
\left( \hat{c}_2\check{c}_2 + \hat{c}_3 \check{c}_3 \right) + f_2
\left(\hat{c}_2 \check{c}_2' + \hat{c}_3 \check{c}_3' + \hat{c}_3'
\check{c}_3\right)\right\}
\\ \nonumber & = \partial_{p} \left\{ -\frac{1}{2} f_1'c_1 + f_2'
\left(
\hat{c}_2\check{c}_2 + \hat{c}_3 \check{c}_3 \right) + f_2
\hat{c}_3' \check{c}_3\right\}
\\ \nonumber & = -\frac{1}{2} f_1''c_1 + f_2''\left(c_2 + c_3\right)
+ f_2' \left(\hat{c}_2 \check{c}_2' + \hat{c}_3 \check{c}_3' +
\hat{c}_3' \check{c}_3 + \hat{c}_3' \check{c}_3\right) + f_2
\left(\hat{c}_3'' \check{c}_3 + \hat{c}_3' \check{c}_3' \right)
\\ & = -\frac{1}{2} f_1''c_1 + f_2''\left(c_2 + c_3\right)
+ 2 f_2' \hat{c}_3' \check{c}_3 + f_2
\left(\hat{c}_3'' \check{c}_3 + \hat{c}_3' \check{c}_3' \right) \hspace{2 pt},
\label{master1}\end{align}
where we used the weak identity $\hat{c}_3 \check{c}_3' = -
\hat{c}_2 \check{c}_2'$ and the convention $\partial_{p} f \equiv f'$ for any 
function $f$. In the above equation there is one 
term proportional 
to $\check{c}_3' \sim \delta((p_\mu +q_\mu)^2)-k^2)$. Already in second 
order derivative expansion this term is problematic as one cannot simply 
interchange the $p\longrightarrow0$ limit with the $q$ integration as this 
would produce terms proportional to $\sim \Theta\left(k^2-q^2\right) 
\times
\delta\left(q^2-k^2\right)$, which is not well defined since the
contribution of the delta function is exactly at the discontinuity of the
Heaviside function. Moreover, subsequent $p$ derivatives will generate 
$\delta'$ terms. The trick in order to deal with these terms is that we 
evaluate the angular integrals for finite $p$, such that there are no more 
$\delta$-distributions involved. Then, we rewrite the terms such that the limit 
$p \longrightarrow 0$ can be taken safely. For the $p^4$ coefficient the 
limit is taken after further $p$-derivatives. More precisely, this trick 
works as follows.  
First we note that all the terms containing $\delta$-functions will be of 
the form
\begin{equation}
\int_{0}^{\infty} 
\int_{-1}^{1} \mathrm{d}q \mathrm{d}x q^3 \sqrt{1-x^2}
\Theta(k^2-q^2) f_2^{(n)}(p,q,x) \hat{c}^{(m)}_3 \check{c}_3'(p,q,x) \,,    
\end{equation}
with the standard notation for the $n$-th derivative with respect to $p$.
We rewrite the $\delta$-function as a function 
of the angular integration variable $x$ according to
\begin{equation}
 \delta\left(f(x)\right) = \delta\left(q^2+p^2+2pqx-k^2\right) =
\frac{1}{2pq}\delta\left(x-\frac{k^2-p^2-q^2}{2pq}\right) \,.
\end{equation} 
Then, we perform the $p$ derivatives acting on the various factors in the 
terms, before killing the angular integration with the $\delta$ function.
After this we are left with terms of the form
\begin{align}\label{int_for_master}
\nonumber & \lim_{p\rightarrow0} \int_{0}^{\infty} 
\int_{-1}^{1} \mathrm{d}q \, \mathrm{d}x \, q^3 \sqrt{1-x^2}  \theta(k^2-q^2)   
\, 
f^{(n)}(p,q,x) \hat{c}^{(m)}_3(p,q,x) \frac{2p+2qx}{2pq} \delta\left(x- 
\frac{k^2-p^2-q^2}{2pq}\right)
\\ \nonumber =& \lim_{p\rightarrow0} \int_{k-p}^{k}\, \mathrm{d}q \,  q^3 
\sqrt{1-\left(\frac{k^2-p^2-q^2}{2pq}\right)^2} 
f^{(n)}\left(p,q,x=\frac{k^2-p^2-q^2}{2pq}\right) 
\\ & \times 
\hat{c}^{(m)}_3\left(p,q,x=\frac{k^2-p^2-q^2}{2pq}\right) \frac{1}{2pq} 
\left(2p +\frac{k^2-p^2-q^2}{p}\right)\,, 
\end{align}
where the new domain of integration of the $q$-integral arises due to the 
condition that the contribution of the $\delta$-function is in the domain of 
integration, which is equivalent to
\begin{equation}
 \frac{k^2-p^2-q^2}{2pq} \in \left(-1,1\right)\,.
\end{equation}
The best way to solve these 
integrals is to keep in mind that we are interested in the limit $p 
\longrightarrow 0$. Therefore, we want to exploit the fact that the domain of 
integration vanishes proportional to $p$ and that all terms of order larger 
than $p^{-1}$ will vanish after integration. Terms of order $p^{-n}$ with $n>1$ 
cannot occur in these expressions. We proceed by writing $q=k - 
\epsilon$ with $\epsilon= y\, p$ and 
transform the integral according to
\begin{equation}\label{integral_transform}
\lim_{p\rightarrow0} \int_{k-p}^{k} \mathrm{d}q \, F(p,q) = - 
\lim_{p\rightarrow0} 
\int_1^0 \mathrm{d}y \, p\, F(p,k-yp) \,.
\end{equation}
Interchanging the limit $p \longrightarrow 0$ with the $y$ integration is now  
trivial and perfectly well defined.
\\Now we can proceed with the derivation of a master formula for the fourth 
momentum-derivative of the flow. Taking two further derivatives of 
\eqref{master1} with respect to $p$ and using $\hat{c}_3 \check{c}_3' = 
-\hat{c}_2 \check{c}_2'$ 
after each derivative, we arrive at
\begin{align}
\nonumber & \partial^{4}_{p}
\left\{-\frac{1}{2}f_1c_1 +
f_2 \left( \hat{c}_2\check{c}_2 + \hat{c}_3 \check{c}_3 \right) \right\}  
\\ \nonumber & = -\frac{1}{2} f_1^{(4)} c_1 + f_2^{(4)}(c_2+c_3) + 
4 f_2^{(3)} \hat{c}'_3 \check{c}_3 + 6 f_2^{(2)} \hat{c}''_3 
\check{c}_3 + 3 f_2^{(2)} \hat{c}'_3 \check{c}'_3 + 4 f_2^{(1)} 
\hat{c}^{(3)}_3 \check{c}_3 
\\ & \,\,\,\,\,\,\,\,\, + 3 f_2^{(1)} \hat{c}^{(2)}_3 \check{c}'_3 + 
f_2 \hat{c}^{(4)}_3 \check{c}_3 +  f_2 \hat{c}^{(3)}_3 
\check{c}'_3  + \partial^2_p \left( f_2 \hat{c}'_3 
\check{c}'_3\right) + 2 \partial_p \left( f_2' \hat{c}'_3 
\check{c}'_3\right) + \partial_p \left( f_2 \hat{c}''_3 
\check{c}'_3\right)  \,.    
\end{align}
In the limit $p \longrightarrow 0$ obviously $\check{c}_3 \longrightarrow 
\theta(q^2-k^2)$. Due to the overall $\theta(k^2-q^2)$, these terms are 
exactly zero except for $q=k$. However, this is just one point, i.e.\ it is a 
domain with zero measure, and the whole integrand is finite at $q=k$.  
Consequently, the limit of vanishing momentum can 
taken before integration and all the terms proportional to $\check{c}_3$  
vanish. In the terms proportional to $c_1$ and $c_2$ the limit is also 
unproblematic, but the contributions are finite. Hence, we are left with
\begin{align}\nonumber
& \int_{0}^{\infty} 
\int_{\mathbb{R}^4} \mathrm{d}q \,\mathrm{d}x \, q^3 \sqrt{1-x^2}  
\theta(k^2-q^2)   
\, 
\left\{-\frac{1}{2}f_1^{(4)}(0,q)c_1+ f_2^{(4)}(0,q)c_2 \right\} + 
\lim_{p\rightarrow0} \partial^2_p \int_{\mathbb{R}^4} \mathrm{d}q \, 
\mathrm{d}x \, 
q^3 \sqrt{1-x^2} f_2 \hat{c}'_3 
\check{c}'_3  \theta(k^2-q^2)  
\\ \nonumber & + 2 \lim_{p\rightarrow0} \partial_p \int_{\mathbb{R}^4} 
\mathrm{d}q \, \mathrm{d}x \, q^3 \sqrt{1-x^2}  \theta(k^2-q^2) f_2' \hat{c}'_3 
\check{c}'_3 + \lim_{p\rightarrow0} \partial_p \int_{\mathbb{R}^4} 
\mathrm{d}q \, \mathrm{d}x \, q^3 \sqrt{1-x^2}  \theta(k^2-q^2) f_2 \hat{c}''_3 
\check{c}'_3
\\ & \lim_{p\rightarrow0} \int_{\mathbb{R}^4} 
\mathrm{d}q \, \mathrm{d}x \, q^3 \sqrt{1-x^2}  \theta(k^2-q^2) \left(3 f_2'' 
\hat{c}'_3 \check{c}'_3 + 3 f_2'\hat{c}''_3 \check{c}'_3 + 
f_2\hat{c}'''_3 \check{c}'_3 \right)\,.            
\end{align}
The terms proportional to $\check{c}'_3$ can be integrated 
according to the above prescription in order to eliminate the 
$\delta$-function, and the residual differentiation with respect to $p$ can 
then be taken afterward, see equations \eqref{int_for_master} and 
\eqref{integral_transform}. As a check, this formula 
was applied to scalar field-theory where we found the correct $p^4$ 
coefficient.

\twocolumngrid

\bibliography{flatgravity}

\end{document}